\documentclass[twocolumn,showpacs,showkeys,superscriptaddress]{revtex4}
\usepackage{graphicx}
\usepackage{bm}
\usepackage{color}
\usepackage{amsmath}
\usepackage{amsfonts}
\usepackage{amssymb}
\usepackage{natbib}
\usepackage{latexsym}
\usepackage{mathrsfs}
\usepackage{slashed}

\begin{document}

\title{Unification of massless field equations solutions for any spin}

\author{Sergio A. Hojman}
\email{sergio.hojman@uai.cl}
\affiliation{Departamento de Ciencias, Facultad de Artes Liberales,
Universidad Adolfo Ib\'a\~nez, Santiago 7491169, Chile.}
\affiliation{Departamento de F\'{\i}sica, Facultad de Ciencias, Universidad de Chile,
Santiago 7800003, Chile.}
\affiliation{Centro de Recursos Educativos Avanzados,
CREA, Santiago 7500018, Chile.}
\author{Felipe A. Asenjo}
\email{felipe.asenjo@uai.cl}
\affiliation{Facultad de Ingenier\'ia y Ciencias,
Universidad Adolfo Ib\'a\~nez, Santiago 7491169, Chile.}

\begin{abstract}
A unification of Klein--Gordon, Dirac, Maxwell, Rarita--Schwinger and Einstein equations exact solutions (for the massless fields cases) is presented. The method is based on writing all of the relevant dynamical fields in terms of products and derivatives of pre--potential functions, which satisfy d'Alambert equation. The coupled equations satisfied by the pre--potentials are non-linear. Remarkably, there are particular solutions of (gradient) orthogonal pre--potentials that satisfy the usual wave equation which may be used to construct {\it{exact non--trivial solutions to Klein--Gordon, Dirac, Maxwell, Rarita--Schwinger and (linearized and full) Einstein equations}}, thus giving rise to a unification of the solutions of all massless field equations for any spin. Some solutions written in terms of orthogonal pre--potentials are presented. Relations of this method to previously developed ones, as well as to other subjects in physics are pointed out. 
\end{abstract}


\maketitle
\section{Introduction}

Klein--Gordon, Dirac, Maxwell, Rarita--Schwinger and Einstein field equations are cornerstones of physics and, as such,  numerous studies have been dedicated on the subject. Many of them are related to solving these equations and it is, therefore, surprising to realize that a unified method to simultaneously produce exact solutions for all of these theories can be devised by introducing pre--potential functions. These are functions which are used to construct massless fields of any spin, and they satisfy d'Alambert equation. Orhogonal pre--potentials, which have gradients which are orthogonal to each other, are extremely useful in this approach.

This method is loosely inspired on two different seemingly unrelated subjects: the search for the two gauge invariant true dynamical degrees of electromagnetism \cite{hoj77} and the solution of the first order inverse problem of the calculus of variations \cite{hojurr}.
It is constructed based on the fact that massless field theories (for non--vanishing spin) have two dynamical degrees of freedom \cite{hoj77}  and that half of Maxwell equations are nothing but the statement that the exterior derivative of a two--form (the electromagnetic field) vanishes. These equations are equivalent to the integrability conditions for Lagrange brackets \cite{hojurr}, where the Lagrange brackets are expressed in terms of gradients of functionally independent constants of motion of the mechanical problem.

The above works as a starting point to construct any spin massless field by using pre--potentials, with the possibility that the same pre--potentials solve all of the equations for different spins.
We first apply the pre-potential method to solve Maxwell equations  exactly and later
apply the same procedure to get solutions for other massless fields of any spin.

\section{Exact solutions of Maxwell equations}

Define the electromagnetic field $F_{ab}(x^c)$ for even--dimensional space-time by
\begin{equation}\label{F2n}
F_{ab}(x^c)=\sum_{i=1}^{n} \left(
\frac{\partial u^{(2i-1)}}{\partial x^a} \frac{\partial u^{(2i)}}{\partial x^b} - \frac{\partial u^{(2i-1)}}{\partial x^b} \frac{\partial u^{(2i)}}{\partial x^a}\right)\ ,
\end{equation}
where the electromagnetic pre--potentials $u^{(c)}(x^b)$ are $2n$ functionally independent real functions of the $2n$ variables $x^b$ with $a, b, c, ....\ =\ 1, 2, 3, ...., 2n$.
 Definition \eqref{F2n} takes advantage of the vanishing exterior derivative of the electromagnetic tensor. Thus, this electromagnetic field satisfies half of the $2n$--dimensional Maxwell equations, $\partial_\gamma F_{\alpha \beta}+ \partial_\beta F_{\gamma \alpha} + \partial_\alpha F_{\beta \gamma}\equiv 0 $, identically ($\partial_\beta\equiv \partial / \partial x^\beta$).
The rest of the (source--free) Maxwell equations 
\begin{equation}\label{Max2}
 \frac{\partial F^{\alpha \beta}}{\partial x^{\beta}} = 0\ .
\end{equation}
define the conditions on the pre-potentials in order to be solutions for the electromagnetic field.

In order to exemplify this, let us now turn our attention to the expression for $F_{\alpha \beta}(x^\gamma)$ written for the $4$--dimensional case
\begin{eqnarray}\label{F4}
F_{\alpha \beta}(x^\gamma)&=&{u^{(1)}}_{,\alpha} {u^{(2)}}_{,\beta}-{u^{(2)}}_{,\alpha} {u^{(1)}}_{,\beta}\nonumber \\
&+&{u^{(3)}}_{,\alpha} {u^{(4)}}_{,\beta}-{u^{(4)}}_{,\alpha} {u^{(3)}}_{,\beta}\ ,
\end{eqnarray}
in a flat Minkowski (pseudo--orthonormal) space--time metric $\eta_{\mu \nu}={\text{diag}}\ (+1, -1, -1, -1)$. Here, the electromagnetic pre-potentials $u^{(\alpha)}(x^\gamma)$ are $4$ functionally independent real functions of $4$ variables $x^\gamma$ with $\alpha, \beta, \gamma, ....\ =\ 0, 1, 2, 3$, and  ${u^{(\alpha)}}_{,\beta}\equiv \partial_\beta u^{(\alpha)}$. 
The usual electromagnetic potential $A_\alpha(x^\beta)$  may  be written in terms of the pre--potentials $u^{(\alpha)}(x^\gamma)$ as
\begin{eqnarray}\label{A1}
A_{\alpha}(x^\gamma)&=&\frac{1}{2}\left({u^{(1)}}_{,\alpha} {u^{(2)}}-{u^{(2)}}_{,\alpha} {u^{(1)}}\right)\nonumber \\
&+&\frac{1}{2}\left({u^{(3)}}_{,\alpha} {u^{(4)}}-{u^{(4)}}_{,\alpha} {u^{(3)}}\right)\nonumber \\
&+& {\Lambda(x^\gamma)}_{,\alpha}\ ,
\end{eqnarray}
where $ {\Lambda(x^\gamma)}$ is an arbitrary function. Of course, all of the pre--potentials $u^{(\gamma)}$ and potentials $A_\alpha$ must be real functions. 

Using the above, 
Eqs.~\eqref{Max2} written explicitly in terms of ${u^{(\alpha)}}$  are
\begin{eqnarray}\label{Max3}
\frac{\partial F^{\alpha \beta}}{\partial x^{\beta}} 
&=&{{u^{(1)}}^{,\alpha}}_{,\beta} {u^{(2)}}^{,\beta}+{u^{(1)}}^{,\alpha} {{u^{(2)}}^{,\beta}}_{,\beta}\nonumber \\
&-&{{u^{(2)}}^{,\alpha}}_{,\beta} {u^{(1)}}^{,\beta}-{u^{(2)}}^{,\alpha} {{u^{(1)}}^{,\beta}}_{,\beta}\nonumber \\
&+&{{u^{(3)}}^{,\alpha}}_{,\beta} {u^{(4)}}^{,\beta}+{u^{(3)}}^{,\alpha} {{u^{(4)}}^{,\beta}}_{,\beta}\nonumber \\
&-&{{u^{(4)}}^{,\alpha}}_{,\beta} {u^{(3)}}^{,\beta}-{u^{(4)}}^{,\alpha} {{u^{(3)}}^{,\beta}}_{,\beta}\ = 0\ .
\end{eqnarray}

It is a straightforward matter to realize that an example of a particular solution in cartesian coordinates is given by
\begin{eqnarray}\label{sol}
u^{(1)}(t,x)&=&p_1(t+x)+p_2(t-x)\ ,\nonumber \\
u^{(2)}(y,z)&=&q_1(y+iz)+{q_1}^*(y-iz)\ ,\nonumber \\
u^{(3)}(t,y)&=&r_1(t+y)+r_2(t-y)\ ,\nonumber \\ 
u^{(4)}(x,z)&=&s_1(z+ix)+{s_1}^*(z-ix) \ ,
\end{eqnarray}
where $p_1$, $p_2$, $r_1$, and $r_2$ are arbitrary real functions and $q_1$ and $s_1$ are arbitrary complex functions.These four pre-potentials are real functions.

In particular, in order to find a general solution of Eqs.~\eqref{Max3}, it is sufficient that
\begin{eqnarray}\label{dal}
 \square  u^{(\gamma)} &=& 0\ , \ \ \forall \gamma \, ,\nonumber\\
{{u^{(2i-1)}}^{,\alpha}}_{,\beta} {u^{(2i)}}^{,\beta} &=& 0\, ,\nonumber\\
{{u^{(2i)}}^{,\alpha}}_{,\beta} {u^{(2i-1)}}^{,\beta}&=& 0\, ,
\end{eqnarray}
for $i=1, 2$,  where $\square$ is the d'Alembert operator in Minkowski space. 
We call pre--potential any function which satisfies d'Alembert equation, and we define orthogonal pre--potentials as any pair of functions which satisfy Eqs.~\eqref{dal}, alluding to the fact that their gradients are orthogonal or equivalently that they, in general, define a (two--dimensional) patch of orthogonal coordinates.

A more general solution, is obtained by requiring that the vectors $v^{(2i-1)}$ and $v^{(2i)}$, defined in terms of the $u^{(\mu)}$ gradients, 
\begin{equation}\label{vec}
{v^{(\mu)}}^{\alpha} = \eta^{\alpha \beta} {u^{(\mu)}}_{,\beta} 
\end{equation}
commute with each other, i.e.,
\begin{equation}\label{comm}
{{v^{(2i-1)}}^{\alpha}}_{,\beta}\ {v^{(2i)}}^{\beta} -\ {{v^{(2i)}}^{\alpha}}_{,\beta}\ {v^{(2i-1)}}^{\beta} = 0\ ,
\end{equation}
for $i =1, 2$.
For Minkowski flat space--time metric, one may require that  the set of coordinates $S^{(\alpha)}(\neq \O , \ \forall\ \alpha)$ on which $u^{(\alpha)}$ depends, fulfill 
\begin{equation}\label{empty}
S^{(1)} \cap\ S^{(2)} =  \O\ \  {\text {and}}\ \  S^{(3)} \cap\ S^{(4)} =  \O\ .
\end{equation}
in order to get an exact particular solution to Maxwell Eqs.~\eqref{Max2}.

 It is also a straightforward matter to prove that the exact electromagnetic field solutions to Maxwell equations given by \eqref{dal} define regular electromagnetic fields  \begin{equation}\label{reg}
 \det  (F_{\alpha \beta})\neq \ 0\ \ ,
\end{equation}
provided the pre--potentials  $u^{(\gamma)}$ are four functionally independent functions, as the ones chosen in example \eqref{sol}, for instance.

\section{Exact solutions of Klein--Gordon equation} 

The wave or Klein--Gordon massless equation reads
\begin{equation}\label{kg}
\square \phi (x^\alpha) =0\ .    
\end{equation}
 Of course, Eq.~\eqref{kg} is solved by any pre--potential.
Nevertheless, in order to achieve a complete unification of all spin fields solutions we would rather choose to write its field solution $\phi (x^\alpha)$ as a product of two orthogonal pre--potentials $u^{(1)}(x^\mu)$ and $u^{(2)}(x^\nu)$,
\begin{equation}
\phi (x^\alpha) = u^{(1)}(x^\mu)\ u^{(2)}(x^\nu)\ ,  
\end{equation}
which is a solution of Klein--Gordon equation \eqref{kg} by properties \eqref{dal}.
In fact, this solution can be generalized to the addition of several product of pairs of orthogonal pre--potentials
\begin{equation}
\phi (x^\alpha) = u^{(1)}(x^\mu)\ u^{(2)}(x^\nu)+u^{(3)}(x^\mu)\ u^{(4)}(x^\nu)+...\, .
\end{equation}

A particular solution can be constructed using pre--potentials \eqref{sol}, impyling that the same pre--potentials solve Maxwell and Klein--Gordon equations.

\section{Exact solutions of Dirac equation}

Consider the massless Dirac equation (Weyl equation)
\begin{equation}\label{mld}
i  \gamma^\mu {\partial}_\mu \psi (x^\alpha)= i\, \slashed\partial  \psi (x^\alpha) =0\, ,
\end{equation}
where Dirac matrices are given the following Kronecker products: $\gamma^0= \sigma^3\otimes I$ and $\gamma^j= i \sigma^2\otimes \sigma^j$, where ${\sigma^j}$ are Pauli matrices and $j=1, 2, 3$.

To solve the massless Dirac equation in a simple way, it is enough to define $\psi (x^\alpha)$ by
\begin{equation}
\psi (x^\alpha)  = \slashed\partial \begin{pmatrix}
 u^{(1)}(x^\mu)\ u^{(2)}(x^\nu)\\
 u^{(3)}(x^\mu)\ u^{(4)}(x^\nu)\\
 u^{(5)}(x^\mu)\ u^{(6)}(x^\nu)\\
 u^{(7)}(x^\mu)\ u^{(8)}(x^\nu)\\
\end{pmatrix} \ ,
\end{equation}  
in terms of pairs of orthogonal pre--potentials. Anew, we can use pre--potentials \eqref{sol} in Dirac equation in order to find a particular solution.

A different exact solution can be constructed with orthogonal pre--potentials. It is known that any solution of the source--free Maxwell equations solves the massless Dirac equation \cite{simulik}. Therefore, any spinor, with  
components $\psi_i (x^\alpha)$ (with $i=1,2,3,4$),
given in terms of orthogonal pre--potentials in the form
\begin{eqnarray}
\psi_1&=&-{u^{(1)}}_{,0} {u^{(2)}}_{,3}+{u^{(2)}}_{,0} {u^{(1)}}_{,3}\nonumber\\
&&
-{u^{(3)}}_{,0} {u^{(4)}}_{,3}+
{u^{(4)}}_{,0} {u^{(3)}}_{,3}\ ,\nonumber\\
\psi_2&=&-{u^{(1)}}_{,0} {u^{(2)}}_{,1}+{u^{(2)}}_{,0} {u^{(1)}}_{,1}\nonumber\\
&&
-{u^{(3)}}_{,0} {u^{(4)}}_{,1}+
{u^{(4)}}_{,0} {u^{(3)}}_{,1}\nonumber\\
&&-i{u^{(1)}}_{,0} {u^{(2)}}_{,2}+i{u^{(2)}}_{,0} {u^{(1)}}_{,2}\nonumber\\
&&
-i{u^{(3)}}_{,0} {u^{(4)}}_{,2}+i
{u^{(4)}}_{,0} {u^{(3)}}_{,2}\, ,\nonumber\\
\psi_3&=&i{u^{(1)}}_{,1} {u^{(2)}}_{,2}-i{u^{(2)}}_{,1} {u^{(1)}}_{,2}\nonumber\\
&&
+i{u^{(3)}}_{,1} {u^{(4)}}_{,2}-i
{u^{(4)}}_{,1} {u^{(3)}}_{,2}\ ,\nonumber\\
\psi_4&=&-{u^{(1)}}_{,3} {u^{(2)}}_{,1}+{u^{(2)}}_{,3} {u^{(1)}}_{,1}\nonumber\\
&&
-{u^{(3)}}_{,3} {u^{(4)}}_{,1}+
{u^{(4)}}_{,3} {u^{(3)}}_{,1}\nonumber\\
&&+i{u^{(1)}}_{,2} {u^{(2)}}_{,3}-i{u^{(2)}}_{,2} {u^{(1)}}_{,3}\nonumber\\
&&
+i{u^{(3)}}_{,2} {u^{(4)}}_{,3}-i
{u^{(4)}}_{,2} {u^{(3)}}_{,3}\, ,
\end{eqnarray}
solves Dirac equation. 

In particular, pre--potentials \eqref{sol} that solve Maxwell and Klein--Gordon equation, also solve Dirac equation. However, other pre--potentials are possible, even complex ones.

\section{Exact solutions of  Rarita--Schwinger equation}

One may write the massless Rarita--Schwinger equations for a vector--spinor $\psi_\beta(x^\alpha)$ as a set of one (Dirac--like) dynamical equation and a couple of constraints \cite{rarita,laurie,zhong,baisya,munczek}, i.e.,
\begin{eqnarray}\label{rs1}
i \slashed\partial \psi_\beta (x^\alpha) &=&0\ ,\nonumber\\
\gamma^\beta \psi_\beta (x^\alpha) &=&0\ ,\nonumber\\
\partial^\beta \psi_\beta (x^\alpha) &=&0\ .
\end{eqnarray}

It is a straightforward matter to prove that the vector--spinor $\psi_\beta (x^\alpha)$, given by
\begin{equation}
\psi_\beta (x^\alpha)  ={\partial}_\beta\left(\slashed\partial u(x^\mu)\right) \slashed\partial \begin{pmatrix}
 u^{(1)}(x^\nu)\\
 u^{(2)}(x^\nu)\\
 u^{(3)}(x^\nu)\\
 u^{(4)}(x^\nu)\\
\end{pmatrix} \ ,
\end{equation}  
solves all of the Eqs. \eqref{rs1} when the pre--potential
 $u(x^\mu)$ is orthogonal to all of the other pre--potentials $u^{(j)}(x^\nu)$ (for $j=1, 2, 3, 4$), a choice which is similar to the one made for Dirac equation. 
 
 For instance, a particular example  for a Rarita--Schwinger field could be given in terms of pre--potentials
\begin{eqnarray}\label{solRS}
u(t,x)&=&p_1(t+x)+p_2(t-x)\ ,\nonumber \\
u^{(1)}(y,z)&=&u^{(3)}(y,z)=q_1(y+iz)+{q_1}^*(y-iz)\ ,\nonumber \\
u^{(2)}(t,y)&=&u^{(4)}(t,y)=p_1(t+x)-p_2(t-x)\ , 
\end{eqnarray}
for arbitrary functions $p_1$, $p_2$, and $q_1$.
All the possible choices for the pre--potentials may coincide with those that solve the equations for spin $0$, $1/2$ and $1$ massless  fields.

\section{Exact solutions of linearized Einstein equations}

Consider linerized Einstein equations for a metric perturbation of Minkowski space--time 
 \begin{equation}\label{g}
g_{\alpha \beta}(x^\gamma)\ =\ \eta_{\alpha \beta}+ h_{\alpha \beta}(x^\gamma)\ ,
\end{equation}
where $\eta_{\alpha\beta}$ is the Minkowski space--time metric and the perturbation $| h_{\alpha \beta}(x^\gamma)| \ll 1$ ($\forall\ \alpha,\beta$).
Linearized Einstein vacuum equations may be written as \cite{MTW}
\begin{equation}\label{linear}
{{{h_\mu}^{\alpha}}_{,\nu \alpha}} + {{{h_\nu}^{\alpha}}_{,\mu \alpha}}-  {{h_{\mu \nu}}_{,\alpha}}^\alpha - h_{,\mu \nu}=0\ ,
\end{equation}
 where $h$ is the trace of perturbed metric $h \equiv {h^\alpha}_\alpha = \eta^{\alpha \beta} h_{\alpha \beta}$.

Inspired in the antisymmetric construction of the electromagnetic field \eqref{F4}, we seek for a solution to Einstein linearized equations by defining a symmetric version of it for $h_{\alpha \beta}$ with gravitational pre--potentials ${ {U}^{(\alpha)}}$, in the form
\begin{eqnarray}\label{h1}
h_{\alpha \beta}(x^\gamma)&=&{{U}^{(1)}}_{,\alpha} { {U}^{(2)}}_{,\beta}+{ {U}^{(2)}}_{,\alpha} { {U}^{(1)}}_{,\beta}\nonumber \\
&+&{{U}^{(3)}}_{,\alpha} { {U}^{(4)}}_{,\beta}+{ {U}^{(4)}}_{,\alpha} { {U}^{(3)}}_{,\beta} \ .
\end{eqnarray}

It is straightforward  to realize that exactly the same electromagnetic orthogonal pre--potentials that solve Maxwell (and Klein--Gordon and Dirac) equations  under the definitions \eqref{dal}, also solve the linearized Einstein equations \eqref{linear} for the {\it{gravitational}} pre--potentials  \eqref{h1}. In order to prove this, first notice that metric \eqref{h1} has
$h=0$ and ${{{h_\mu}^{\alpha}}_{,\alpha}}=0$,  by conditions \eqref{dal} and \eqref{comm}. Of course, a coordinate (gauge) transformation may always be introduced in expression \eqref{h1}. Thereby, the equations for the gravitational pre--potentials ${ {U}^{(\alpha)}}$ are obtained using Eq.~\eqref{linear} to get
\begin{eqnarray}
&&\sum_{i=1}^{i=2}\left({ {U}^{(2i-1)}}_{,\mu ,\nu}\ \square { {U}^{(2i)}} + { {U}^{(2i)}}_{,\mu ,\nu}\ \square { {U}^{(2i-1)}}\right)\nonumber \\
&&=\sum_{i=1}^{i=2} \left({ {U}^{(2i-1)}}_{,\mu ,\alpha} {{{ {U}^{(2i)}}}^{,\alpha}}_{,\nu}+{ {U}^{(2i-1)}}_{,\nu ,\alpha} {{{ {U}^{(2i)}}}^{,\alpha}}_{,\mu}\right)\nonumber \\    
&&=0\ ,\end{eqnarray}
 which are identically satisfied if conditions \eqref{dal} are met.

It is important to stress that these solutions are non--trivial as long as they produce a non--identically vanishing Riemann tensor to first order in the smallness parameter.

\section{Exact solutions of full Einstein equations}

It is remarkable that some of the solutions to the linearized Einstein theory {\it{also satisfy exactly the full theory  in vacuum}}, with no approximations whatsoever. Exact spacetime metrics can be constructed using the pre--potentials. In these cases, the metric have the form
\begin{equation}
g_{\alpha\beta}={\hat g}_{\alpha\beta}+\Theta_{\alpha\beta}\, ,
\end{equation}
 where ${\hat g}_{\alpha\beta}$ is a base flat metric, and
now $\Theta_{\alpha\beta}$ is not a perturbation, but it has the same form than Eq.~\eqref{h1}, i.e., 
$\Theta_{\alpha \beta}(x^\gamma)={{U}^{(1)}}_{,\alpha} { {U}^{(2)}}_{,\beta}+{ {U}^{(2)}}_{,\alpha} { {U}^{(1)}}_{,\beta}+{{U}^{(3)}}_{,\alpha} { {U}^{(4)}}_{,\beta}+{ {U}^{(4)}}_{,\alpha} { {U}^{(3)}}_{,\beta}$. For the case of full Einstein equations, the pre--potencials and their derivatives are not small, in general.

We can explicitly write some exact metrics that solve the full Einstein equations in terms of pre--potentials. In cartesian coordinates, a metric that solve the system \cite{codemathematica} is written for the base Minkowski metric ${\hat g}_{\alpha\beta}=\eta_{\alpha\beta}$, and
\begin{eqnarray}
U^{(1)}(x,t)&=&\xi_1(x+t)\, ,\nonumber\\
U^{(2)}(y,z)&=&\xi_2(y+i z)+\xi_2(y-i z)\, ,\nonumber\\
U^{(3)}(x,t)&=&\xi_3(x+t)\, ,\nonumber\\
U^{(4)}(y,z)&=&\xi_4(y+i z)+\xi_4(y-i z)\, ,
\end{eqnarray}
where $\xi_i$ (with $i=1,2,3,4$) are arbitrary functions. Notice that this exact spacetime is not, in general, a wave. Besides, its Riemann tensor is not identically zero, in general, and thus is a non--flat space--time solution. Besides, this solution can be generalized to introduce free parameters in it. For example, the pre--potentials
\begin{eqnarray}
U^{(1)}(x,t)&=&\xi_1(x+t)\, ,\nonumber\\
U^{(2)}(y,z)&=&\xi_2\left(e^{-i\alpha}(y+i z)\right)+\xi_2\left(e^{i\alpha}(y-i z)\right)\, ,\nonumber\\
U^{(3)}(x,t)&=&\xi_3(y-t)\, ,\nonumber\\
U^{(4)}(y,z)&=&\xi_4(x+i z)+\xi_4(x-i z)\, ,
\end{eqnarray}
also solve full Einstein equations in vacuum, where again  $\xi_i$ are arbitrary functions, and $\alpha$ is an arbitrary constant. Other generalizations are possible.

For  a cylindrical form of the flat metric, with  coordinates ($t,r,\theta,z$) and base metric 
${\hat g}_{00}=-1=-{\hat g}_{rr}=-{\hat g}_{zz}$, and ${\hat g}_{\theta\theta}=r^2$ (all other components vanish), an exact solution of Eistein equations is found when 
\begin{eqnarray}
U^{(1)}(z,t)&=&\zeta(z-t)\, ,\nonumber\\
U^{(2)}(r)&=&\ln r\, , 
\end{eqnarray}
where $\zeta$ is an arbitrary function, and $U^{(3)}=0=U^{(4)}$. This spacetime metric gives rise to a non--vanishing Riemann tensor (and it does not represent a  wave, in general).

There are also solutions for a base metric with  the light--like form of the flat metric  for coordinates ($u,v,y,z$), given by ${\hat g}_{uv}=1={\hat g}_{yy}={\hat g}_{zz}$ (all other components vanish). For this case, it is enough to consider
\begin{eqnarray}
U^{(1)}(u)&=&\chi_1(u)\, ,\nonumber\\
U^{(2)}(y,z)&=&\chi_2(y+i z)+\chi_2(y-i z)\, ,
\end{eqnarray}
while $U^{(3)}=0=U^{(4)}$. Again, $\chi_i$ (with $i=1,2$) are arbitrary functions.
This spacetime has anew a Riemann tensor that is not identically zero.

Finally, other exact non--trivial solutions can be obtained for 
a light--like cylindrical form of the flat metric, for coordinates ($u, v, r, \theta$), with
${\hat g}_{uv}=1$, ${\hat g}_{rr}=1$, and ${\hat g}_{\theta\theta}=r^2$, and other vanishing components.
In this case, the pre--potencials read
\begin{eqnarray}
U^{(1)}(r)&=&\rho_1(r)\, ,\nonumber\\
U^{(2)}(r,\theta)&=&A \cosh \left(m \ln r\right)  \sin\left(m\, \theta\right)\, ,\nonumber\\
U^{(3)}(v)&=&\rho_2(v)\, ,\nonumber\\
U^{(4)}(r,\theta)&=&A \cosh \left(w \ln r\right)  \sin\left(w\, \theta\right)\, ,
\end{eqnarray}
where $A$, $m$ and $w$ are arbitrary constants, and $\rho_i$ ($i=1,2$) are arbitrary functions.

A clarifying comment seems to be in order. All the orthogonal pre--potentials may be used to construct (massless) solutions to Klein--Gordon, Dirac, Maxwell, Rarita--Schwinger and the linearized Einstein equations. Some of them even solve the full Einstein theory exactly. In the case of (linearized or full) gravity one should check that the Riemann tensor is not identically zero (to have a non--flat space--time solution) for the full theory and up to first order in the linearized case. The above solutions have such property.

\section{Discussion}

The approach we present may be described, loosely speaking, as a unified way of constructing solutions for massless field equations for any spin where the fields are made up of a core and a shell. The core is common to all fields of any spin and is made up of (exactly the same) orthogonal pre--potentials. The shells depend upon the spin of the fields. It is important to remark that bosonic and fermionic fields share exactly the same core. It is worth noting that linearized and full gravity share both the core and the shell. The most remarkable fact is that one can find examples (which we have presented) where exactly the same core pre--potentials solve the massless Klein--Gordon, Dirac, Maxwell, Rarita--Schwinger and linearized and full Einstein equations, which shows a deep connection between different spin fields. The approach may lead to a different way of understanding supersymmetry (keeping the core fixed while changing the shell). On the other hand, the orthogonal pre--potentials may, in principle, be used to determine the algebraic structure of the fields.

There are several other approaches which have a similar (but different) ways to deal with either different spin fields and/or solutions to field equations. Among them, Feynman and Gell--Mann dealt with a kind of pre--potential for the Dirac equation \cite{fgm} and Penrose \cite{pen69} presented a kind of pre--potential which once integrated differs from spin to spin making it less transparent to relate solutions of different spin fields among them. There are also many other early and recent methods devised by Clebsch \cite{clebsch}, Bateman \cite{bat}, Geroch \cite{geroch}, A\c cik \cite{acik}, and Bia\l inicky--Birula \cite{bb2021}, to deal with solutions to field equations, for instance.

 Our approach may also be used to deal with topological aspects of fields as Ra\~nada \cite{ranada} has done. Furthermore, it is also possible to extend the pre--potential method to find dynamical solutions for higher-spin fields, which follows from extended Maxwell-like equations in the massless case  \cite{dkross}.
Finally, is important to  stress that the current approach allows us to construct exact solutions to the full non--linear Einstein equations. This possibility will be explored further in forthcoming articles.



\begin{thebibliography}{17}

\bibitem{hoj77} S. Hojman, Ann. of Phys. (N.Y.), {\bf 103}, 74 (1977).
\bibitem{hojurr} S.A. Hojman and L.F. Urrutia, J. Math. Phys. {\bf 22}, 1897 (1981).
\bibitem{simulik} V. M. Simulik  and I. Yu. Krivsky, Proceedings, 2nd International Conference on Symmetry in Nonlinear Mathematical Physics, V. 2, 475 (1997).
\bibitem{rarita} W. Rarita and J. Schwinger, Phys. Rev. {\bf 60}, 61 (1941).
\bibitem{laurie} D. Luri\'e, {\it Particles and Fields} (John Wiley and Sons, Inc. 1968).

\bibitem{zhong} H. Shi-Zhong, R. Tu-Nan, W. Ning and Z. Zhi-Peng, Eur. Phys. J. C {\bf 26}, 609  (2003).



\bibitem{baisya} H. L Baisya, Nucl. Phys. {\bf B29}, 104 (1971).
\bibitem{munczek} H. Munczek, Phys. Rev. {\bf 164}, 1794 (1967).





\bibitem{MTW} C. W. Misner, K. S. Thorne and J. A. Wheeler, {\it Gravitation} (W. H. Freeman and Company, San Francisco, 1971).



\bibitem{codemathematica} All computations in linearized and full Einstein theory can be performed using the ${\it{Mathematica}} \textsuperscript{\textregistered}$ code ``Curvature and the Einstein Equation'' written by Leonard Parker, \url{http://web.physics.ucsb.edu/~gravitybook/math/curvature.pdf}. 

\bibitem{fgm} R.P. Feynman and M. Gell--Mann, Phys. Rev, {\bf 109}, 193, (1958)
\bibitem{pen69} R. Penrose, J. Math. Phys., {\bf 10}, 38, (1969)

\bibitem{clebsch} A. Clebsch(1859), Journal f\"ur die Reine und Angewandte Mathematik,  {\bf 56} 10, (1859)
\bibitem{bat} H. Bateman, {\it{ ``The Mathematical Analysis of Electrical and
Optical Wave-Motion on the Basis of Maxwell‘s Equations''}}, (Cambridge University Press, Cambridge, 1915) p. 12.
\bibitem{geroch} R. Geroch, J. Math. Phys. {\bf 12}, 918 (1971); {\bf 13}, 394 (1972)
\bibitem{acik} \"O. A\c cik and \"U. Ertem, Phys. Rev. D {\bf 98}, 066004 (2018).
\bibitem{bb2021} I. Bialynicki-Birula, arXiv:2101.03325 (2021).
\bibitem{ranada} A. F. Ra\~ nada, Lett. Math. Phys. {\bf 18}, 97 (1989).


\bibitem{dkross} D. K. Ross, Nuovo Cim. {\bf 58}, 11 (1980).


\end{thebibliography}
\end{document}